\documentclass[sigconf]{acmart}
\AtBeginDocument{%
  }

\setcopyright{acmlicensed}
\copyrightyear{2018}
\acmYear{2018}
\acmDOI{XXXXXXX.XXXXXXX}
\acmConference[Conference acronym 'XX]{Make sure to enter the correct
  conference title from your rights confirmation email}{June 03--05,
  2018}{Woodstock, NY}
\acmISBN{978-1-4503-XXXX-X/2018/06}

\copyrightyear{2026}
\acmYear{2026}
\setcopyright{cc}
\setcctype{by}
\acmConference[CHI EA '26]{Extended Abstracts of the 2026 CHI Conference on Human Factors in Computing Systems}{April 13--17, 2026}{Barcelona, Spain}
\acmBooktitle{Extended Abstracts of the 2026 CHI Conference on Human Factors in Computing Systems (CHI EA '26), April 13--17, 2026, Barcelona, Spain}
\acmDOI{10.1145/3772363.3798943}
\acmISBN{979-8-4007-2281-3/2026/04}

\begin{document}

\title{MicroVRide: Exploring 4-in-1 Virtual Reality Micromobility Simulator}


\author{Xiaoyan Zhou}
\orcid{0000-0002-1650-8314}
\affiliation{%
  \institution{KTH Royal Institute of Technology}
  \city{Stockholm}
  \country{Sweden}}
\email{xizhou@kth.se}

\author{Natalia Sempere}
\orcid{0009-0003-5336-4860}
\affiliation{%
  \institution{KTH Royal Institute of Technology}
  \city{Stockholm}
  \country{Sweden}}
\email{sempere@kth.se}

\author{Pooria Ghavamian}
\orcid{0009-0009-8146-2744}
\affiliation{%
  \institution{KTH Royal Institute of Technology}
  \city{Stockholm}
  \country{Sweden}}
\email{pooriag@kth.se}

\author{Asreen Rostami}
\orcid{0000-0003-0594-3027}
\affiliation{%
  \institution{RISE Research Institutes of Sweden}
    \city{Stockholm}
  \country{Sweden}}
\affiliation{%
  \institution{Stockholm University}
  \city{Stockholm}
  \country{Sweden}}
\email{asreen.rostami@ri.se}

\author{Andrii Matviienko}
\orcid{0000-0002-6571-0623}
\affiliation{%
  \institution{KTH Royal Institute of Technology}
  \city{Stockholm}
  \country{Sweden}}
\email{andriim@kth.se}

\renewcommand{\shortauthors}{Zhou et al.}

\begin{abstract}
Micromobility vehicles, such as e-scooters, Segways, skateboards, and unicycles, are increasingly adopted for short-distance travel due to their low weight and low emissions. Despite their growing popularity, we lack controlled, low-risk environments to study rider experiences and performance. While virtual reality (VR) simulators offer a promising approach by reducing safety risks and providing immersive experiences, micromobility simulators remain largely underexplored. We introduce MicroVRide, a modular 4-in-1 VR micromobility simulator that supports e-scooters, Segways, electric unicycles, and one-wheeled skateboards on a single platform. The simulator preserves vehicle-specific physical constraints and control metaphors, enabling the study of diverse riding behaviors with minimal hardware reconfiguration. We contribute the simulator design and report a preliminary within-subject study (N = 12) that demonstrates feasibility and reveals distinct experiential profiles across vehicles.
\end{abstract}



\begin{teaserfigure}
\centering
  \includegraphics[width=\textwidth]{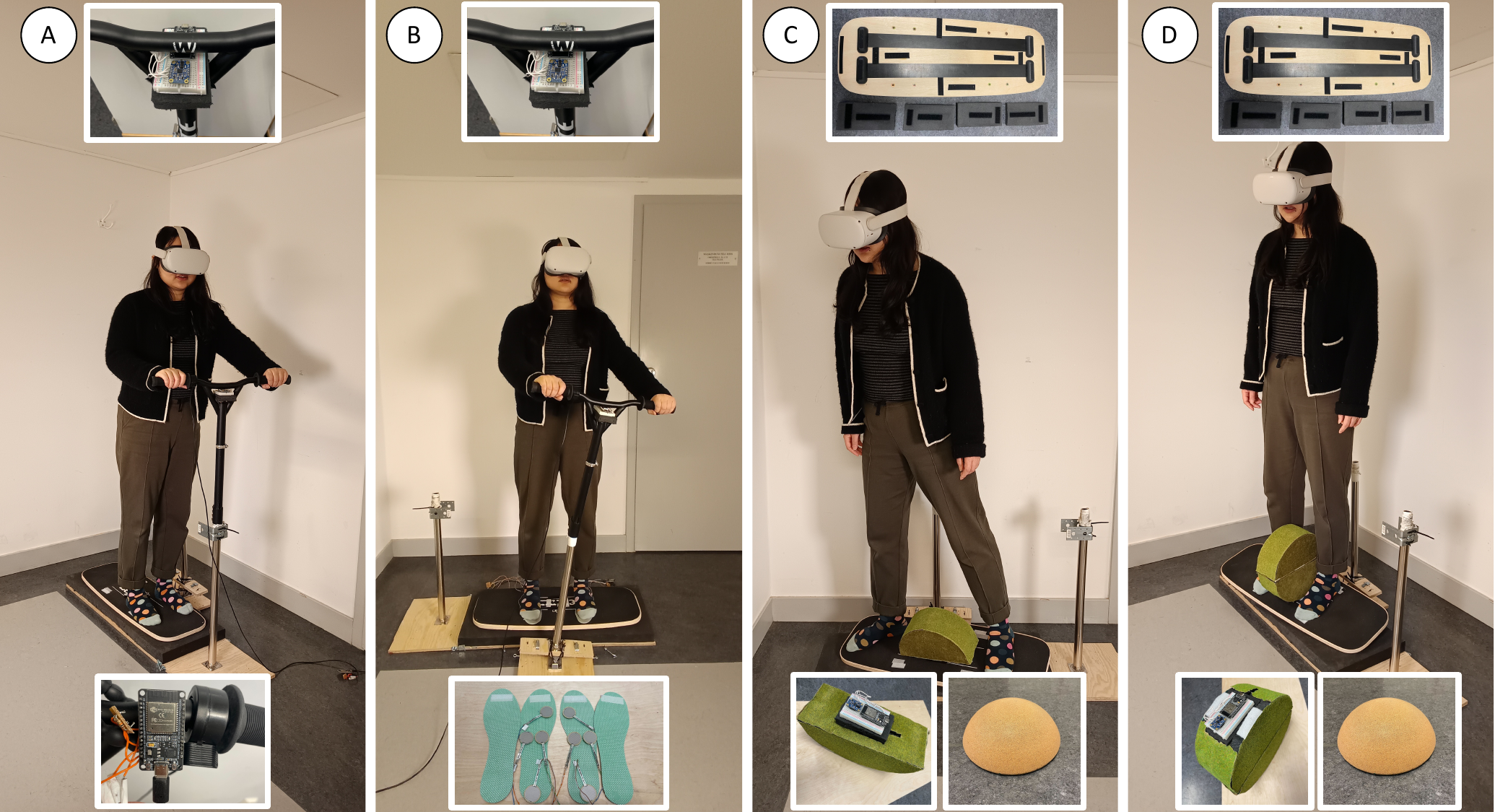}
  \caption{MicroVRide 4-in-1 micromobility VR simulator. 
(A) \textbf{E-scooter:} handlebar with IMU (yaw) $\rightarrow$ steering; thumb throttle $\rightarrow$ velocity. 
(B) \textbf{Segway:} handlebar with IMU (roll) $\rightarrow$ steering; force-sensitive resistors insoles $\rightarrow$ velocity.
(C) \textbf{One-wheeled skateboard:} a half-wheel prop between user’s feet; hemispherical balance base for omni-directional rotation; IMU maps platform pitch/roll $\rightarrow$ velocity/steering; soft buffers damp impacts.
(D) \textbf{Electric unicycle:} full-wheel prop between user’s feet; hemispherical balance base for omni-directional rotation; IMU maps platform pitch/yaw $\rightarrow$ velocity/steering; soft buffers damp impacts.}
  \Description{}
  \label{fig:teaser}
\end{teaserfigure}


\maketitle

\section{Introduction}
Micromobility has rapidly expanded as an everyday transportation mode, offering flexible and low-emission alternatives to car-based travel~\cite{ITF_SafeMicromobility, WSJ2024MicromobilityStreetSafety, BUNING2025100164, MicromobilityAdoption2024}. Vehicles such as e-scooters, Segways, electric unicycles, and one-wheeled skateboards differ in steering, speed regulation, and bodily coordination \cite{BarcelonaSunSegwayBeginnersGuide, RiderGuideHowToRideScooter, InMotionHowToRideEUC, OnewheelWikiRidingTechnique}. Studying such behavior in real-world traffic is costly and potentially unsafe~\cite{SafeInfrastructure2023, MicromobilitySafety2024}, particularly with multiple micromobility vehicles or novice riders.
VR driving simulators provide controllable, repeatable environments for training and safety evaluation~\cite{DriveSimulatorSchool2018, DrivingSimulatorGaze2022, DriveSimulatorTraining2017, DrivingSimulatorEntertainment2017, kataria2023virtual}, and have been applied to micromobility for behavior and interaction studies \cite{RollMotion2024, E-bikeSimulator2021, MicrovehiclesSimulators2025}.
Beyond transportation-focused studies, recent work has also explored micromobility-inspired simulators as embodied locomotion interfaces for virtual navigation, demonstrating their potential for broader VR interaction scenarios beyond mobility research, such as immersive navigation and spatial exploration \cite{NAVIS2024, LocoScooter2026}.

Bicycle simulators constitute most of established research in micromobility simulation by instrumenting bicycles mounted on fixed or motion-enabled rigs to capture steering input, pedaling behavior, and rider posture, in combination with immersive virtual environments to study user interaction and safety-relevant scenarios \cite{RollMotion2024,MicroMobility2025,CyclingSimulator2025,MicrovehiclesSimulators2025, E-bikeSimulator2021, NovelEBicycleSimulator2025, Wintersberger2022, Matviienko2023CyclingFidelity, MatviienkoVRSickness2022, Keppel2026, 10.1145/3229434.3229479, 10.1145/3706598.3713407}. 
Although e-scooter~\cite{ZouSim2021, Lefeuvre, Matviienko2022} and handlebar-free Segway~\cite{Zhao2023} simulators have recently gained attention alongside the bicycle simulators, VR simulators for electric unicycles and one-wheeled skateboards remain largely underexplored. 
Moreover, most existing simulators are designed for a single vehicle type and rely on fixed control mappings~\cite{ZouSim2021, Lefeuvre, MicroMobility2025, E-bikeSimulator2021}, and research on a broader set of micromobility vehicles with varying steering mechanisms and body-based control strategies remains limited. 
These limitations motivate the need for VR micromobility simulators that can support multiple vehicle configurations within a shared experimental platform.

In this work, we present MicroVRide, a modular VR micromobility simulator that supports four vehicle configurations: e-scooters, Segways, electric unicycles, and one-wheeled skateboards within a single physical and software platform. MicroVRide preserves vehicle-specific steering and control metaphors while allowing researchers to study each vehicle using a shared experimental setup. We describe the system design and preliminary findings from a within-subject study (N = 12) that examines the feasibility and user experience of interacting with micromobility vehicles in VR. Our work contributes a reusable research platform and early design considerations for VR-based micromobility simulation.

\section{MicroVRide}
MicroVRide is a modular VR micromobility simulator that supports multiple vehicles on a shared physical and software infrastructure. The system combines a reconfigurable physical platform with multiple sensor modalities and a Unity application with a unified data pipeline, enabling rapid switching between different micromobility vehicles while preserving vehicle-specific control characteristics, and reconfiguration took approximately one minute.
MicroVRide captures users’ body-based input through motion and force sensors integrated into the riding platform. Sensor data are transmitted wirelessly to map physical input to virtual vehicle motion (Figure~\ref{fig:teaser}).
The physical platform is designed for standing interaction and supports interchangeable configurations across all vehicles. 
For vehicles that require a stable standing surface (e-scooter and Segway), the platform remains fixed. For vehicles that rely on full-body input (electric unicycles and one-wheeled skateboards), a hemispherical balance element is placed beneath the platform to enable continuous 360-degree rotation. A detachable handlebar can be quickly and easily added or removed depending on the vehicle configuration.
Inertial measurement units (IMUs) capture rotational input (pitch, roll, yaw) and are mounted either on the handlebar (e-scooter, Segway) or directly on the platform (electric unicycle, one-wheeled skateboard). Force-sensitive resistors embedded in insoles are used in the Segway to capture foot-pressure input for forward and backward motion, reflecting real-world control \cite{BarcelonaSunSegwayBeginnersGuide}. A thumb throttle on the handlebar provides speed control in the e-scooter. Across all vehicles, sensor data is streamed via Wi-Fi to the Unity application running on the VR HMD.

\begin{figure*}
    \centering
    \includegraphics[width=\linewidth]{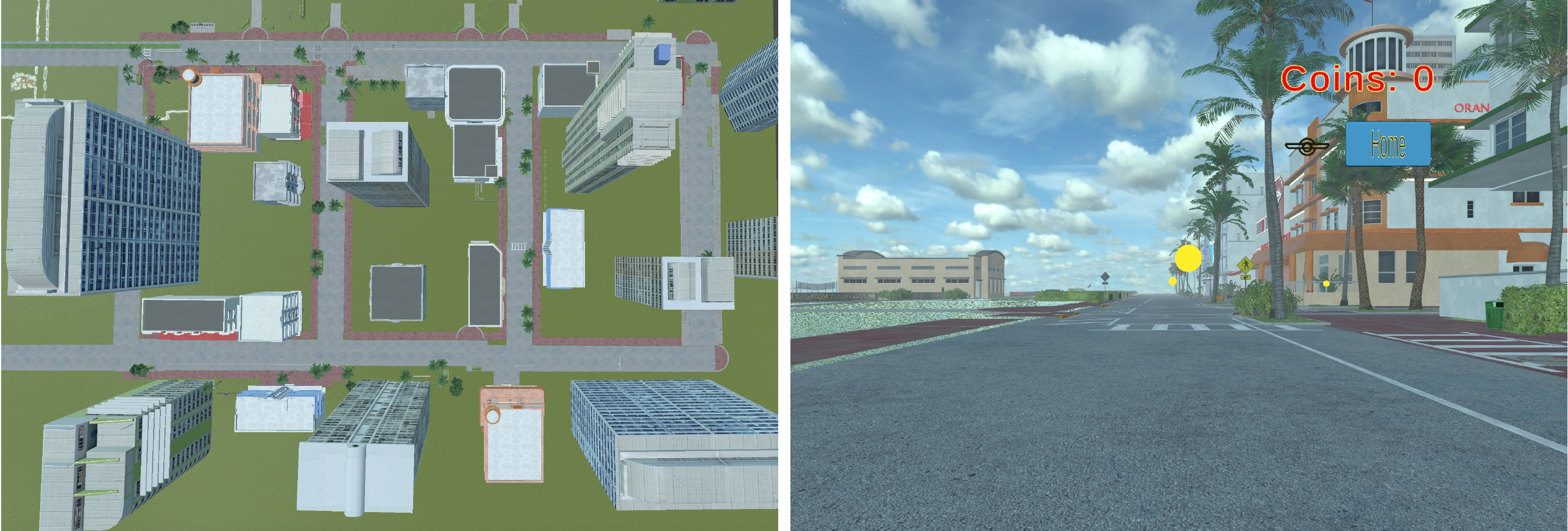}
    \caption{We investigated four routes  with each vehicle (left) that contained coins placed at equal distances from each other (right).}
    \label{fig:simulation}
\end{figure*}





Control mappings reflect existing real-world interaction patterns. For example, the e-scooter uses handlebar-based steering combined with throttle-controlled speed. The Segway maps foot-pressure input to forward and backward motion and uses handlebar roll for steering, while the standing platform remains fixed. The unicycle and one-wheeled skateboard rely entirely on IMUs, mapping body rotation to speed and steering. Although these two configurations share a similar physical platform, the IMU is mounted and referenced differently, resulting in distinct control mappings that better match the interaction characteristics of each vehicle. All control mappings are implemented as modular software components that can be activated based on the selected vehicle in VR.
The virtual simulation (Figure~\ref{fig:simulation}) is implemented in Unity and integrates vehicle physics, environment rendering, and interaction logic. A centralized data receiver manages incoming sensor streams and distributes normalized input values to the active vehicle controller. Each vehicle is implemented as a dedicated controller module that encapsulates its control logic, motion constraints, and parameterization, while sharing a common interface for input and state updates. This design allows new vehicle types or alternative control strategies to be introduced by extending controller modules without changes to the sensing or networking layers.
MicroVRide includes a logging system that records time-stamped sensor values, normalized control inputs, vehicle state variables, and discrete system events. 

IMU orientation estimates are computed using onboard sensor fusion (accelerometer, gyroscope, magnetometer) to derive pitch, roll, and yaw angles. Orientation values are filtered to reduce jitter before being mapped to vehicle control signals, and axis alignment is calibrated per vehicle configuration to match the physical mounting orientation with the virtual reference frame. In the Segway configuration, force-sensitive resistor (FSR) values are normalized per participant. Baseline values are recorded with no load on the sensors, followed by a maximum reference measurement in which participants stand on the fixed board and apply strong downward pressure. Each FSR signal is normalized between these bounds, and a small dead zone prevents unintended motion due to sensor noise. Although formal end-to-end latency measurements were not conducted, no noticeable delay between physical input and visual vehicle response was reported during the study. The hemispherical element is fixed beneath the platform, while a balance board placed on top enables user-driven omni-directional tilt and rotation. The mechanism is fully passive, with mechanical limits and damping elements restricting excessive tilt for safety.

\section{Preliminary Study}
We did not intend to provide a comparative performance evaluation between vehicle types. Instead, we validated that multiple micromobility vehicles can be experienced within a single simulator setup and identified early design considerations for future simulator iterations.
In the study, participants wore a VR HMD. They interacted with the physical simulator platform configured for four micromobility vehicles: an e-scooter, a Segway, an electric unicycle, and a one-wheeled skateboard. Each participant experienced all configurations in a single session. The order of vehicles was counterbalanced to mitigate ordering effects. For each vehicle, participants first completed a short training phase, followed by a coin-collecting task in a simple virtual city environment designed to encourage maneuverability (Figure~\ref{fig:simulation}). The four routes were structurally comparable, with similar path length and coin spacing across configurations. Trials were self-paced and lasted until participants completed the route or until participants chose to stop due to discomfort. Training duration was not fixed and continued until participants reported feeling comfortable with the controls. Participants had the option to hold onto a wall for balance support during interaction. After completing each condition, they filled out a post-condition questionnaire. The questionnaire included a raw NASA-TLX to capture perceived workload, and a post-condition questionnaire with five questions focused on control confidence, perceived stability and predictability, and presence and embodiment. After experiencing all four vehicles, participants provided qualitative feedback on their experience of the simulator.

\begin{figure*}
    \centering
    \includegraphics[width=\linewidth]{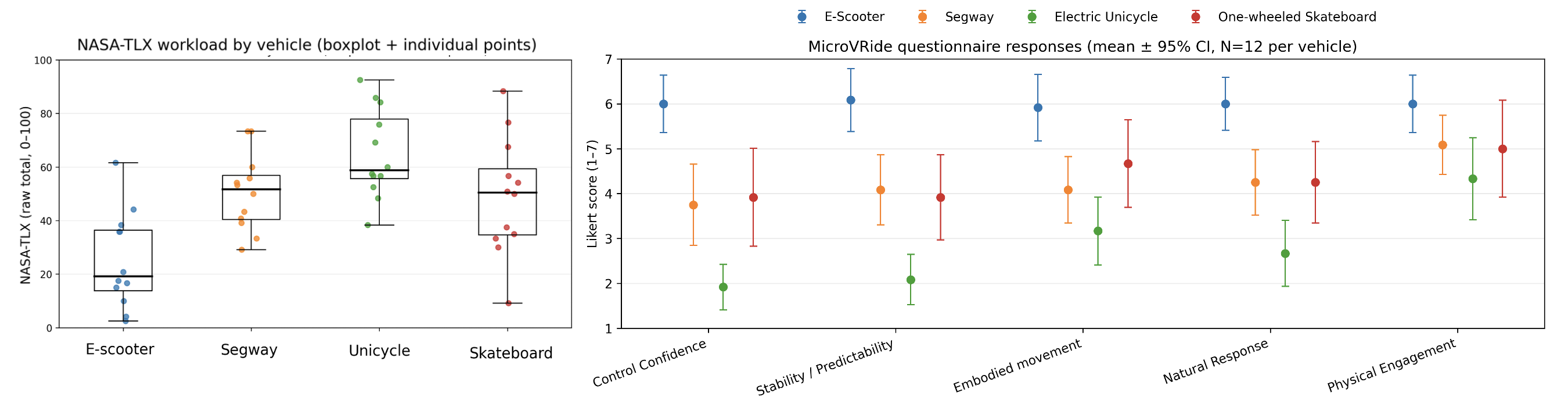}
    \caption{(Left) Raw NASA-TLX scores per vehicle. (Right) Likert ratings for control, stability/predictability, and embodiment.}
    \label{fig:Questionnaire}
\end{figure*}

\begin{figure*}
    \centering
    \includegraphics[width= 0.7\linewidth]{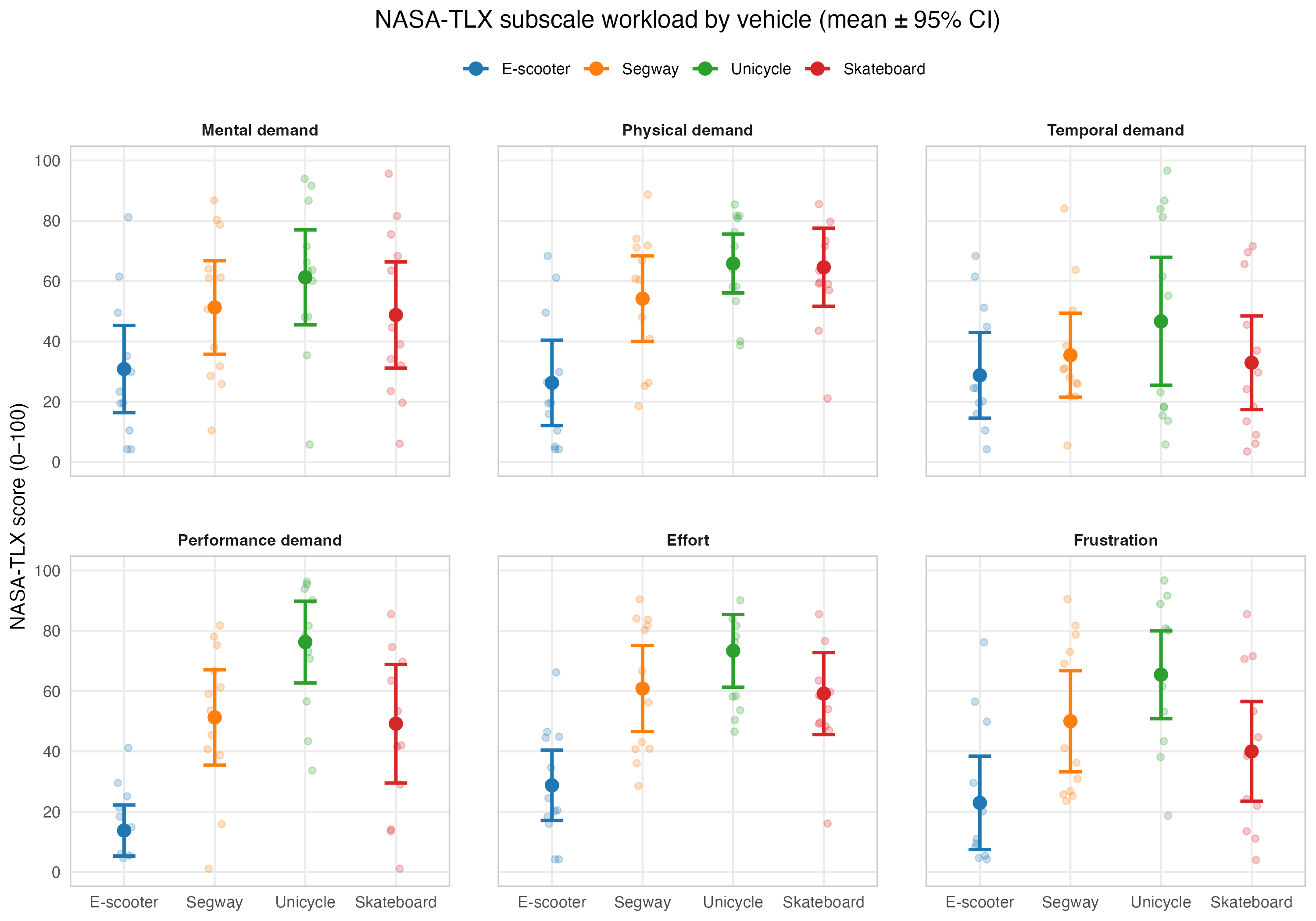}
    \caption{NASA-TLX subscale scores by vehicle (mean ± 95\% CI). Higher scores indicate greater perceived workload.}
    \label{fig:subscales}
\end{figure*}

\section{Results}
We recruited twelve participants (9M, 2F, 1 non-binary): 3 aged 18–24, 8 aged 25–34, 1 aged 35–44. VR experience varied: 3 participants used VR frequently, 6 -- occasionally, 2 -- a few times, and 1 had no prior experience. 10 participants reported prior experience riding an e-scooter, 1 reported prior experience riding a Segway, and 1 reported no experience with any of the four target vehicles. Three participants reported experience with skateboards or wave boards, and two -- with balance-based game interfaces, e.g., Wii Fit. Most participants reported no (N=4) or only mild (N=8) susceptibility to motion sickness. Two participants withdrew after the first condition due to motion sickness and were replaced. Given the preliminary nature of this study and the limited sample size (N = 12), we report descriptive statistics to characterize observed trends rather than conducting inferential comparisons.

\textbf{Workload}. Raw NASA-TLX showed differences in perceived workload across the four configurations. Participants reported the lowest workload in the e-scooter condition ($M = 25.21, SD = 17.26$), moderate workload in the Segway ($M = 50.47, SD = 20.05$) and skateboard ($M = 49.96, SD = 15.54$), and the highest workload in the electric unicycle ($M = 64.77, SD = 16.29$) (Figure ~\ref{fig:Questionnaire}). Figure ~\ref{fig:subscales} provides a breakdown of NASA-TLX subscales, illustrating that workload differences were driven by distinct combinations of demands across vehicle types. The e-scooter showed low demand ratings, with the lowest perceived performance pressure ($13.8$). In contrast, the Segway produced elevated workload, especially in Effort ($60.8$), Physical Demand ($54.2$), and Mental Demand ($51$.2), suggesting that participants experienced the Segway mapping as requiring sustained embodied Effort and active control monitoring. The electric unicycle showed the highest workload across nearly all TLX dimensions, with high scores in Performance demand ($76.2$), Effort ($73.3$), and Frustration ($65.4$), indicating that participants frequently perceived the unicycle condition as difficult to control and demanding to stabilize. While the skateboard had a comparable level of Physical Demand ($64.6$) to the unicycle ($65.8$), it resulted in substantially lower Frustration ($40.0$), suggesting that participants experienced it as physically engaging but less cognitively stressful.

\textbf{Control, Stability, and Embodiment}.
SegwayControl confidence (\textit{“I felt in control of the vehicle”}) and stability/predictability (\textit{“The vehicle felt stable and predictable”}) were highest for the e-scooter and lowest for the unicycle (Figure ~\ref{fig:Questionnaire}). This aligns with the TLX findings, which showed that the e-scooter produced the lowest workload and the unicycle the highest.
Participants also reported the strongest perceived coupling between their body movements and vehicle motion while riding the unicycle, indicating that the board configuration supported a strong sense of embodied control despite requiring substantial physical engagement. The e-scooter received the highest ratings for naturalness (\textit{“The vehicle’s responses felt natural given what I was doing physically”}), consistent with its direct and familiar throttle-and-handlebar control metaphor. Finally, participants reported the greatest physical engagement with the Segway, suggesting that its foot-pressure-driven speed control encouraged active bodily involvement even when perceived workload remained moderate. 

\textbf{Interview}.
Participants reported that the four MicroVRide configurations were distinguishable and produced unique experiences. They emphasized variability across vehicles (e.g., "very different from vehicle to vehicle", "different vehicles gave me different experiences."), suggesting that the simulator successfully conveyed distinct control rather than feeling like a generic VR locomotion system.
Participants described the simulator as engaging and enjoyable, highlighting the immersive environment and the novelty of embodied interaction: \textit{"I'm impressed with the, the feel like the city view and, Yeah, the whole experience."} [P12]) and \textit{"It's actually easier to ride on VR compared to the real world, because it's easier to balance, because it's like flat. And in the real world, it's like with the wheels, it's kind of wonky." }[P8]
Although participants described higher physical demand for certain vehicles, \textbf{effort} was framed as a meaningful part of the experience: \textit{"different vehicles have different, like effort to move."}[P6] and \textit{"Some of them were way more physically demanding and not as responsive as others. While the remaining was really engaging (e-scooter and Segway), really fun, and something I would like to do again." } [P11].
Participants highlighted bodily engagement as a core feature of MicroVRide: \textit{"I mean, it's interesting, especially that it would get your like physical body engaged."} [P4] and \textit{"It's like it's challenging, but it's fun to try." }[P6].
They also valued the \textbf{tangible physical feedback} from the platform compared to purely virtual interactions: \textit{"I liked that there were like actual motions... since that's static, you don't really get any like perceived feedback from, from the actual hardware. So I like that with you, you could feel the balancing and feel that things were happening." } [P9].

Participants' \textbf{perceived difficulty} aligned with the workload and control. The e-scooter was consistently described as easiest (\textit{"The easiest, of course., was the scooter. The input was very fast. And you don't have to do any balancing…"}[P7], and \textit{"Scooter. The easiest indeed."}[P11]) and the unicycle as hardest (\textit{"I think the hardest one was probably the unicycle. Because of the balancing platform itself."}[P7] and \textit{"And the hardest was my first one (unicycle)."}[P11]).
Participants underlined the importance of a handlebar for controllability: \textit{"I felt like the ones with handles, for me at least, were very much easier to control than the ones without...So then, like a handle just felt more natural."}[P3], and \textit{"the one without a handle control was quite hard to control." }[P5] 
At the same time, they mentioned strong controllability on configurations without a handlebar: \textit{"the scooter and the one like the skateboard, basically, those two were, like, very fun. I felt I had full control of the vehicles"}[P12], or \textit{"the skateboard felt quite more or less easy to control." }[P7] 
This variation suggests that perceived controllability is influenced not only by vehicle configuration but also by individual factors such as prior experience and bodily control strategies.
\textbf{Prior experience} with micromobility appeared to shape how intuitive the mappings felt. In particular, the e-scooter was frequently described as realistic: \textit{"The e-scooter was very natural. It felt just like riding a normal one. The steering was realistic."}[P7], \textit{"Well, I'm used to riding a scooter, so that was somewhat familiar."}[P5], and \textit{"I think especially the e-scooter one, as it translates very well in my head." }[P9]
Participants also attributed difficulty with the unicycle/skateboard to limited real-world familiarity: \textit{"Maybe it's because I don't have, like prior experience in using like the balancing vehicles (unicycle and skateboard). So it's rather difficult overall."}[P6], and \textit{"I'm bad at the vehicles I'm normally bad at, I don't ride them because I know my balance is bad." }[P8] 
This suggests that experience may shape perceived workload and learning demands during transition to embodied VR control.

Participants emphasized lower-body engagement as the primary strategy in configurations without handlebars (\textit{"My feet like lower body, I feel like pretty much entirely...I feel like basically all the work went into like the lower body, legs and feet."}[P3]) and whole-body strategy for unicycle control (\textit{"With unicycle, you have to use your whole body, your arms to balance."}[P7] and \textit{"for the unicycle, My legs and like my hips."}[P3]).
Participants reported low \textbf{motion sickness}, consistent with the background questionnaire that showed slight or no motion sickness: \textit{"I didn't feel sick at all."}[P1] and \textit{"I didn't feel dizzy or sick at any point." }[P7] 
Participants who reported transient discomfort mentioned that it was quickly resolved: \textit{"I felt it (dizziness), and then it, like, dropped away." }[P10]
Discomfort was often tied to discrete failure events rather than continuous locomotion. 
Multiple participants linked dizziness to camera rotation when colliding with obstacles: \textit{"I think, for the motion sickness, I... feel most... when I stumbled into something"}[P6], \textit{"only the times where I went too fast and hit an obstacle in VR, it will start to flip"}[P11], and \textit{"when I fell, everything Just like, rotate. Like the view."}[P12] This suggests that micro-events (e.g., sudden camera rotation) may be more influential for discomfort than steady-state riding.
Participants' \textbf{suggestions} focused primarily on responsiveness and system feedback, such as limitations in acceleration/deceleration: \textit{"I had a little bit of a problem with the acceleration and deceleration. I felt like maybe it wasn't so responsive." } [P1] 
Several participants questioned whether falling animations were necessary given their disruptive effect on comfort: \textit{"The thing is, like, when I bump into, like, the sidewalks and then it has a fail animation. So I think it's, it's not that necessary to have the following animation." }[P8] 
One participant highlighted that the mapping between feet movements and the vehicle control was unclear due to limited visual feedback of lower-body input \textit{"I feel like you can't see your feet, like if it's moving the right way, or am I putting so much pressure in the wrong direction? So it's difficult because you can't see your feet. Maybe, maybe, seeing an arrow when I turn, because sometimes I don't know if I'm turning right or left. It would help me to learn how to control the simulator faster." } [P2]

\section{Discussion and Future Work}
A key observation from our study concerns cybersickness and individual differences in tolerance, which are crucial for VR locomotion research. While most participants reported little or no motion sickness, two participants withdrew after the first condition due to strong discomfort, indicating meaningful variability in susceptibility even within a small sample. This withdrawal underscores the need for adaptive safety protocols in embodied VR locomotion systems, particularly given known individual differences in cybersickness tolerance. Participants' descriptions further suggest that discomfort may be linked less to steady forward motion and more to abrupt camera changes, especially when the vehicle flipped or collided. This aligns with the known sensitivity of vestibular–visual mismatch during sudden viewpoint rotations \cite{VRSickness2020, MotionSickness2006}. These findings indicate that future simulator iterations should prioritize more conservative failure animations, particularly when the simulator is used for longer study sessions or less experienced users.
Moreover, the results indicate that MicroVRide can facilitate early-stage skill acquisition and adaptation, even for participants with limited real-world micromobility experience. Most participants reported prior experience primarily with e-scooters, yet many engaged with the Segway, electric unicycle, and one-wheeled skateboard configurations after short practice. Participants also described transferring knowledge from related prior experiences, such as skateboarding or balance-based interfaces, to interpret the simulator's control mappings and adapt their movement strategies. While our study was not designed to evaluate training effectiveness, VR-based micromobility simulation may provide a promising environment for controlled exposure to unfamiliar vehicle control styles.
Participants' feedback also highlighted the importance of embodiment and visual feedback in the learning process. 
This suggests that future iterations could benefit from real-time feedback overlays to support learning and reduce uncertainty, especially during early familiarization.
Finally, our findings emphasize practical considerations for deploying a reusable simulator across diverse participants. In particular, hardware adjustability, such as handlebar height for different body sizes and adaptable placement of the pressure-sensing insoles, appears important to ensure usability and comfort across users. 
This highlights that, beyond sensing and control fidelity, factors such as stability, perceived robustness, and clear affordances may strongly shape users' willingness to commit to full-body motion in VR. Improving the physical build quality and visual appearance of the simulator contributes not only to safety but also to perceived reliability and overall experience.

\begin{acks}
This work was supported by the Digital Futures project MicroVRide.
\end{acks}

\bibliographystyle{ACM-Reference-Format}
\bibliography{sample-base}

\appendix

\end{document}